\documentclass[preprint,12pt]{elsarticle}

\usepackage{amssymb}
\usepackage{amsmath}
\usepackage{xcolor}
\usepackage{graphicx}
\usepackage{subfigure}

\begin{document}

\begin{frontmatter}



\title{Influence of intraspecies interactions on the nucleation and wetting phase diagram in dilute ternary Bose-Einstein condensates}


\author{Nguyen Van Thu} 

\affiliation{organization={Department of Physics, Hanoi Pedagogical University 2},
            city={Hanoi},
            postcode={100000}, 
            country={Vietnam}}

\begin{abstract}
Within the framework of Gross–Pitaevskii theory, we investigate the effects of intraspecies interactions on the nucleation transition and the wetting phase diagram of dilute ternary Bose–Einstein condensate in the regime of strong segregation between two components. The analyses are carried out using both the analytical double-parabola approximation (DPA) and numerical computations. Our results show that the DPA provides a reliable approximation for describing the nucleation transition. For the wetting phase diagram, we find that the DPA is in excellent agreement with numerical results in symmetric systems, particularly in the completely symmetric case, whereas it fails to provide an adequate description for asymmetric systems.
\end{abstract}


%
%
%


\end{frontmatter}



\section{Introduction}

The wetting phenomenon is a well-known and widely observed occurrence in both natural and everyday contexts \cite{Gennes1985,Bonn2009}. In classical fluid systems, wettability is typically analyzed in terms of intermolecular interactions. In the context of quantum physics, within the Ginzburg-Landau theory of type-I superconducting materials \cite{1965}, also surface phase transitions of wetting type were discovered \cite{Indekeu1995}. Within the framework of mean-field theory for critical phenomena, the wetting phase transition in superconductors is interpreted as the delocalization or unbinding of the interface between the normal and superconducting phases in type-I superconductors \cite{Indekeu1995}. This wetting phase transition has been experimentally observed in conventional superconductors \cite{Kozhevnikov2007}. Furthermore, the phenomenon of wetting phase transitions has been extensively studied in magnetic systems \cite{Hu2023}.

The wetting phenomenon in Bose-Einstein condensates (BECs) was first predicted theoretically in 2004 \cite{Indekeu2004}, in the context of a two-component BECs confined by an idealized hard-wall (optical wall). Despite the theoretical interest it has garnered, this phenomenon has not yet been observed experimentally, even after more than two decades. Several hypotheses have been advanced to explain this absence, accompanied by a range of theoretical and experimental efforts aimed at overcoming the associated challenges. One key limitation is the ideal nature of the hard-wall potential, which cannot be realized in practical experimental setups. To address this issue, alternative approaches have been proposed, including the replacement of the hard-wall confinement with a more experimentally accessible soft-wall potential \cite{VanSchaeybroeck2015}, and investigations into the thermodynamic behavior of the prewetting transition \cite{DuyThanh2024}.

A more recent development involves the substitution of the conventional hard- or soft-wall boundary by a third constituent within a ternary (three-component) BECs system \cite{Indekeu2025}. This configuration is characterized not only by the atomic masses $m_i, (i=1,2,3)$, but also by six intrinsic interaction parameters: three intraspecies scattering lengths $a_{ii}$, and three interspecies scattering lengths $a_{ij}=a_{ji}, (i,j=1,2,3, i\neq j)$. Additionally, the system comprises three independent particle densities $n_i$. The presence of a third component in BEC systems markedly increases the number of experimentally controllable parameters, particularly the relative coupling constants $K_{13},K_{23}$ and the healing length ratios $\xi_i/\xi_j$ (see below). These parameters can be tuned either independently or in conjunction, in BECs mixtures composed of two hyperfine states of a single atomic species or of two distinct atomic species \cite{McCarron2011,Egorov2013,Bauer2009,Liu2014,Kanjilal2023}. A notable example involves laser-induced shifts of magnetic Feshbach resonances in $^{87}$Rb, which permit precise tuning of the scattering lengths while maintaining significantly lower particle loss rates than those associated with optical Feshbach resonances \cite{Bauer2009}. Furthermore, multicomponent systems featuring both ferromagnetic and antiferromagnetic spin–spin interactions offer an extended degree of control \cite{Kanjilal2023}. In mixtures of $^{87}$Rb and $^{133}$Cs, the interspecies scattering length, which characterizes atomic interactions, has been successfully manipulated using magnetic Feshbach resonances in magnetic quadrupole traps, followed by confinement in levitated optical traps \cite{McCarron2011}, or through the use of cigar-shaped trapping geometries \cite{Egorov2013}. This enhanced tunability gives rise to increasingly rich and intricate wetting phase diagrams, underscoring the complexity and flexibility of multicomponent BEC systems.

In Ref. \cite{Indekeu2025}, the wetting phase diagram was systematically analyzed within the parameter space spanned by the interspecies interaction coupling constants, while the intraspecies interaction strengths were held fixed. Phase diagrams were constructed for both symmetric and asymmetric configurations, revealing the existence of degenerate points. Both first-order and critical wetting transitions were identified within these regimes. Recently, the wetting phase diagram within these regimes was investigated by numerical computation for strong segregation between components 1 and 2 and intermediate segregation of components in case of the symmetry and antisymmetry of the intrinsic atomic parameters \cite{Berx2026}. In the regime of strong segregation between components 1 and 2, analytical relation for the nucleation line and numerical computations for the first-order and critical wetting lines showed that the wetting phase transition is the degenerate first-order, i.e., these three lines coincide in the phase diagram formed by the relative coupling constants. This fact leads to conclusion that the DPA is not suitable in this regime. As the remaining focus of this work, we investigate the nucleation phase transition and wetting phase diagram of  dilute ternary BEC in the parameter space defined by the ratios of healing lengths within both the DPA and GP theory combining with numerical computation. These ratios are experimentally tunable by varying the intraspecies atomic interaction strengths while keeping the interspecies interactions fixed.

The paper is organized as follows: Section \ref{sec:wavefunction} we have a brief the DPA for the wave functions of a ternary BEC in strong segregation limit between components 1 and 2. The nucleation phase transition and wetting phase diagram are investigated in Section \ref{wettingphasediagram}.  Conclusions are reported in Section \ref{Conclusions}.

\section{Wave functions of dilute ternary BEC in double parabola approximation\label{sec:wavefunction}}

In this Section we summarize the results for wave functions in DPA presented in Ref. \cite{Indekeu2025}. We start by considering dilute ternary BEC consisting of three components confined within a volume $V$ in the absence of external fields. Within the framework of the grand canonical ensemble, the thermodynamic properties of the system are governed by the grand potential functional
\begin{eqnarray}
\Omega=\int_V d^3\vec{r}\left\{\sum_{i=1}^3\left[-\frac{\hbar^2}{2m_i}\psi_i^*(\vec{r})\nabla^2\psi_i(\vec{r})\right]+V_{\rm GP}\right\},\label{eq:Omega}
\end{eqnarray}
in which Gross-Pitaevskii (GP) potential is
\begin{eqnarray}
V_{\rm GP}=\sum_{i=1}^3\left[-\mu_i|\psi_i(\vec{r})|^2+\frac{G_{ii}}{2} |\psi_i(\vec{r})|^4\right]+ \sum_{i < j} G_{ij} |\psi_i(\vec{r})|^2 |\psi_j(\vec{r})|^2.\label{VGP}
\end{eqnarray}
Here $\hbar$ denotes the reduced Planck constant, $\mu_i$ is the chemical potential of the $i$-th component, and $\psi_i(\vec{r})$ is the macroscopic wave function associated with that component. In the absence of particle flow, the wave functions can be taken as real and serve as order parameters. The local condensate density of each component is then defined by $n_i (\vec{r}) \equiv |\psi_i(\vec{r}) |^2 $. The interaction strengths are characterized by the coupling constants. In the dilute BECs, where the $s$-wave scattering lengths $a_{ii}$ and $a_{ij}$ are much smaller than the average interatomic spacing, the interatomic interactions can be modeled as contact potentials. In this limit, the intraspecies coupling constants take the form
\begin{eqnarray}
G_{ii}=\frac{4\pi\hbar^2a_{ii}}{m_i},\label{gii}
\end{eqnarray}
while the interspecies coupling constants are given by
\begin{eqnarray}
G_{ij}=2\pi\hbar^2\left(\frac{1}{m_i}+\frac{1}{m_j}\right)a_{ij},\label{aij}
\end{eqnarray}
For convenience, we introduce the dimensionless interspecies coupling parameters
\begin{eqnarray}
K_{ij}\equiv\frac{G_{ij}}{\sqrt{G_{ii}G_{jj}}}=\frac{m_i+m_j}{2\sqrt{m_im_j}}\frac{a_{ij}}{\sqrt{a_{ii}a_{jj}}}.\label{K}
\end{eqnarray}
which quantify the relative strength of interspecies interactions. In this work, we restrict our attention to the immiscible regime, characterized by $K_{ij}>1$. For each component, the characteristic length is the healing length, which is defined as $\xi_j\equiv \hbar/\sqrt{2m_i\mu_i}=\hbar/\sqrt{2m_jG_{ii}n_i}$. Therefore, the healing length ratio depends on atomic parameter \cite{VanSchaeybroeck2015,Indekeu2025}
\begin{eqnarray}
\frac{\xi_i}{\xi_j}=\left(\frac{n_ja_{jj}}{n_ia_{ii}}\right)^{1/2}=\left(\frac{m_ja_{jj}}{m_ia_{ii}}\right)^{1/4}.\label{xiratio}
\end{eqnarray}

We now proceed to describe the system under consideration. The system exhibits translational symmetry in the $(x,y)$-plane and is inhomogeneous along the $z$-axis. At the initial time, two condensates, labeled as components 1 and 2, are in two-phase equilibrium, and a stable interface between them, referred to as the 1–2 interface, is present. This implies that the bulk pressures of the two condensates are equal $P_1=P_2\equiv P$, where the bulk pressure $P_i$ of component $i$ is given by
\begin{eqnarray}
P_i=\frac{\mu_i^2}{2G_{ii}}.\label{pressure}
\end{eqnarray} 
Condensate 1 occupies the bulk region at $z\rightarrow -\infty$ while condensate 2 occupies the bulk region at $z\rightarrow \infty$. Subsequently, a third component (condensate 3) is introduced at the 1–2 interface and is considered a candidate for the wetting phase. For convenience, we define the auxiliary chemical potential $\bar\mu_3$, the auxiliary condensate density $\bar n_3$ and the auxiliary healing length $\bar\xi_3$,  which satisfy the following relations
\begin{eqnarray}
\bar\mu_3&=&\sqrt{\frac{G_{33}}{G_{11}}}\mu_1=\sqrt{\frac{G_{33}}{G_{22}}}\mu_2,\nonumber\\
\bar n_3&=&\frac{\bar\mu_3}{G_{33}},\nonumber\\
\bar\xi_3&=&\frac{\hbar}{\sqrt{2m_3\bar\mu_3}}.\label{auxiliary}
\end{eqnarray}
It is evident from these definitions that $\mu_3\leq\bar \mu_3$ and $n_3\leq\bar n_3$. Under this configuration, the boundary conditions for the condensate wavefunctions in the bulk are given by
\begin{eqnarray}
&&\psi_1(-\infty)=\sqrt{n_1},~\psi_2(\infty)=\sqrt{n_2},\nonumber\\
&&\psi_1(\infty)=\psi_2(-\infty)=\psi_3(\infty)=\psi_3(-\infty)=0.\label{boundaries}
\end{eqnarray}
By performing rescalings analogous to those introduced in Ref. \cite{Indekeu2025}, namely, $\psi_1=\sqrt{n_1}\tilde\psi_1, \psi_2=\sqrt{n_2}\tilde\psi_2, \psi_3=\sqrt{\bar n_3}\tilde\psi_3$ and $z=\xi_2\tilde z$ one obtains a set of three coupled GP equations, as detailed in Ref. \cite{Indekeu2025}
\begin{eqnarray}
\label{coupledGP}
   \left (\frac{\xi_1}{\xi_2}\right )^2    \frac{d^2\tilde \psi_1}{d\tilde z ^2}  &=& - \tilde  \psi_1 + \tilde \psi_1^3 + \Sigma_{j \neq 1} \,K_{1j} \,\tilde \psi_j^2 \,\tilde \psi_1, 
    \nonumber \\
    \frac{d^2\tilde \psi_2}{d\tilde z ^2} &=& - \tilde  \psi_2 + \tilde \psi_2^3 + \Sigma_{j \neq 2} \,K_{2j}\, \tilde \psi_j^2 \,\tilde \psi_2, 
    \nonumber \\
    \left (\frac{\bar\xi_3}{\xi_2}\right )^2    \frac{d^2\tilde \psi_3}{d\tilde z ^2} &=& - \frac{\mu_3}{\bar \mu_3}\tilde  \psi_3 + \tilde \psi_3^3 + \Sigma_{j \neq 3} \,K_{3j} \,\tilde \psi_j^2 \,\tilde \psi_3.\,\, 
\end{eqnarray}
The boundary conditions (\ref{boundaries}) reduce
\begin{eqnarray}
&&\tilde\psi_1(-\infty)=1,~\tilde\psi_2(\infty)=1,\nonumber\\
&&\tilde\psi_1(\infty)=\tilde\psi_2(-\infty)=\tilde\psi_3(\infty)=\tilde\psi_3(-\infty)=0.\label{boundaries1}
\end{eqnarray}

In the remainder of this work, we focus on the regime in which components 1 and 2 are in the strong segregation limit, characterized by $K_{12}\rightarrow\infty$ as treated within the framework of the DPA introduced in Ref. \cite{Indekeu2015}. We choose the coordinate origin $\tilde z=0$ to coincide with the interface between condensates 1 and 2. Under this limit, the GP equations (\ref{coupledGP}) are rewritten as
\begin{eqnarray}
\label{coupledGP1}
   \left (\frac{\xi_1}{\xi_2}\right )^2    \frac{d^2\tilde \psi_1}{d\tilde z ^2}  &=& - \tilde  \psi_1 + \tilde \psi_1^3 + K_{13} \,\tilde \psi_3^2 \,\tilde \psi_1, 
    \nonumber \\
    \tilde \psi_2&=&0, 
    \nonumber \\
    \left (\frac{\bar\xi_3}{\xi_2}\right )^2    \frac{d^2\tilde \psi_3}{d\tilde z ^2} &=& - \frac{\mu_3}{\bar \mu_3}\tilde  \psi_3 + \tilde \psi_3^3 + \,K_{13} \,\tilde \psi_1^2 \,\tilde \psi_3\,\, 
\end{eqnarray}
for $\tilde z\leq 0$ and
\begin{eqnarray}
\label{coupledGP2}
   \tilde \psi_1&=&0, 
    \nonumber \\
    \frac{d^2\tilde \psi_2}{d\tilde z ^2} &=& - \tilde  \psi_2 + \tilde \psi_2^3 + K_{23}\, \tilde \psi_3^2 \,\tilde \psi_2, 
    \nonumber \\
    \left (\frac{\bar\xi_3}{\xi_2}\right )^2    \frac{d^2\tilde \psi_3}{d\tilde z ^2} &=& - \frac{\mu_3}{\bar \mu_3}\tilde  \psi_3 + \tilde \psi_3^3 + K_{23} \,\tilde \psi_2^2 \,\tilde \psi_3,\,\, 
\end{eqnarray}
for $\tilde z\geq 0$. The boundary conditions (\ref{boundaries1}) become
\begin{eqnarray}
&&\tilde\psi_1(-\infty)=1,~\tilde\psi_2(\infty)=1,\nonumber\\
&&\tilde\psi_1(\tilde z\geq 0)=\tilde\psi_2(\tilde z\leq0)=\tilde\psi_3(\infty)=\tilde\psi_3(-\infty)=0.\label{boundaries2}
\end{eqnarray}

the system is naturally partitioned into three spatial domains. Domain I, defined by $\tilde z\in (-\infty,\tilde z^-]$, comprises condensates 1 and 3, whose density profiles intersect at a point $\tilde z= \tilde z^-$. In this region, the GP potential (\ref{VGP}) is expanded to second order in deviations from the bulk density of component 1 (i.e., $\tilde\psi_1=1$) and around zero for component 3. The resulting DPA potential in domain I takes the form
\begin{eqnarray}
\widetilde{V}_{\rm I}^{\rm(DPA)}=-2(1-\tilde\psi_1)^2-\left(K_{13}-\frac{\mu_3}{\bar\mu_3}\right)\tilde\psi_3^2.\label{DPAI}
\end{eqnarray}
Here we use $\widetilde{V}\equiv V/(g_{11}n_1^2/2)$, which is dimensionless potential. From (\ref{DPAI}) the coupled GP equations in (\ref{coupledGP}) become
\begin{eqnarray}
\left(\frac{\xi_1}{\xi_2}\right)^2\tilde\psi_1''&=&-2(1-\tilde\psi_1),\nonumber\\
\left(\frac{\bar\xi_3}{\xi_2}\right)^2\tilde\psi_3''&=&\left(K_{13}-\frac{\mu_3}{\bar\mu_3}\right)\tilde\psi_3.\label{GPI}
\end{eqnarray}
The general solutions to Eq. (\ref{GPI}), subject to the boundary conditions specified in Eq. (\ref{boundaries1}), are given by
\begin{eqnarray}
\tilde\psi_1=1-A_1e^{\frac{\xi_2}{\xi_1}\tilde z},\tilde\psi_2=0,~\tilde\psi_3=A_3e^{\sqrt{K_{13}-\frac{\mu_3}{\bar\mu_3}}\frac{\xi_2}{\bar\xi_3}\tilde z}.\label{solution1}
\end{eqnarray}
Domain III, defined by $\tilde z\in [\tilde z^+,\infty)$, contains condensates 2 and 3, which intersect at $\tilde z= \tilde z^+$. By symmetry with domain I, the DPA potential corresponding solutions in domain III is
\begin{eqnarray}
\widetilde{V}_{\rm III}^{\rm(DPA)}=-2(1-\tilde\psi_2)^2-\left(K_{23}-\frac{\mu_3}{\bar\mu_3}\right)\tilde\psi_3^2,\label{DPAIII}
\end{eqnarray}
therefore the wave functions for the condensates are
\begin{eqnarray}
\tilde\psi_1=0,~\tilde\psi_2=1-D_2e^{-\sqrt{2}\tilde z},~ \tilde\psi_3=D_3e^{-\sqrt{K_{23}-\frac{\mu_3}{\bar\mu_3}}\frac{\xi_2}{\bar\xi_3}\tilde z}.\label{solution3}
\end{eqnarray}
Domain II, defined by $\tilde z\in[\tilde z^-,\tilde z^+]$ serves as the intermediate region, in which condensates 1 and 3 coexist for $\tilde z\in [\tilde z^-,0]$ and condensates 2 and 3 coexist for $\tilde z\in [0,\tilde z^+]$. In this domain, the wave function of condensate 3 varies from its maximum value $\sqrt{\mu_3/\bar\mu_3}$ whereas the deviation of wave functions for condensates 1 and 2 are small from zero. The DPA potential reads
\begin{eqnarray}
\label{DPAII}
    \widetilde{V}^{(\rm DPA)}_{\rm II} =
    &-&\frac{1}{2} + \frac{1}{2}\left (\frac{\mu_3}{\bar \mu_3}\right )^2
    -2 \frac{\mu_3}{\bar \mu_3}\left(\sqrt{\frac{\mu_3}{\bar \mu_3}}-\tilde \psi_3\right)^2  \nonumber \\&-&\left(\frac{\mu_3}{\bar \mu_3}K_{13}-1\right) \tilde \psi_1^2 
    -\left(\frac{\mu_3}{\bar \mu_3}K_{23} - 1\right)\tilde \psi_2^2. 
\end{eqnarray}
The condensate wavefunctions in this region take the form
\begin{eqnarray}
\label{solution2}
\tilde \psi_1  &=&   2 B_1 \,\sinh \left(\sqrt{\frac{\mu_3}{\bar \mu_3}K_{13}-1} \,\frac{\xi_2}{\xi_1}\tilde z \right), \; \mbox{for} \; \tilde z < 0, \nonumber
    \\
    \tilde \psi_2  &=&   -2C_2 \,\sinh \left(\sqrt{\frac{\mu_3}{\bar \mu_3}K_{23}-1} \,\tilde z \right), \; \mbox{for} \; \tilde z > 0, \nonumber
        \\
  \tilde \psi_3 &=& \sqrt{\frac{\mu_3}{\bar \mu_3}} + B_3 \,\exp{\left(\sqrt{\frac{2\mu_3}{\bar \mu_3}} \frac{\xi_2}{\bar\xi_3}\tilde z \right)}+ C_3\, \exp{\left(- \sqrt{\frac{2\mu_3}{\bar \mu_3}} \frac{\xi_2}{\bar\xi_3}\tilde z \right)}.
   \end{eqnarray} 
The general solutions in Eqs. (\ref{solution1}), (\ref{solution3}) and (\ref{solution2}) contain eight integration constants $A_1,A_3, B_1, B_3, C_2,C_3,D_2,D_3$. These constants are determined by enforcing continuity of the condensate wave functions and their first derivatives. Specifically, continuity conditions for $\tilde\psi_1,\tilde\psi_3$  are imposed at $\tilde z^-$ while those for $\tilde\psi_2,\tilde\psi_3$ are imposed at $\tilde z^+$, 
\begin{eqnarray}
\tilde\psi_1(\tilde z\rightarrow\tilde z^-+0)=\tilde\psi_1(\tilde z\rightarrow\tilde z^--0),~\tilde\psi_3(\tilde z\rightarrow\tilde z^-+0)=\tilde\psi_3(\tilde z\rightarrow\tilde z^--0),\nonumber\\
\tilde\psi_2(\tilde z\rightarrow\tilde z^++0)=\tilde\psi_2(\tilde z\rightarrow\tilde z^+-0),~\tilde\psi_3(\tilde z\rightarrow\tilde z^++0)=\tilde\psi_3(\tilde z\rightarrow\tilde z^+-0),\nonumber\\
\tilde\psi_1'(\tilde z\rightarrow\tilde z^-+0)=\tilde\psi_1'(\tilde z\rightarrow\tilde z^--0),~\tilde\psi_3'(\tilde z\rightarrow\tilde z^-+0)=\tilde\psi_3'(\tilde z\rightarrow\tilde z^--0),\nonumber\\
\tilde\psi_2'(\tilde z\rightarrow\tilde z^++0)=\tilde\psi_2'(\tilde z\rightarrow\tilde z^+-0),~\tilde\psi_3'(\tilde z\rightarrow\tilde z^++0)=\tilde\psi_3'(\tilde z\rightarrow\tilde z^+-0).\label{BC}
\end{eqnarray}
Although this procedure yields a complete system of eight equations, the resulting expressions are algebraically lengthy and do not yield additional physical insight. For this reason, we omit their explicit forms here.

\section{Nucleation and wetting phase diagram and loci of degenerate points\label{wettingphasediagram}}

In this section, we examine the nucleation and wetting phase transition and derive the corresponding wetting phase diagram. To this end, we begin by recalling some results previously reported in Ref. \cite{Indekeu2025}. As discussed in the preceding section, we initially consider a system comprising two components, labeled 1 and 2, which coexist in a two-phase equilibrium. Under these conditions, the interface between components 1 and 2 represents a stable configuration. Subsequently, component 3 is introduced at the 1-2 interface. This third component acts as a surfactant under the condition
\begin{eqnarray}
P_3=\left(\frac{\mu_3}{\bar\mu_3}\right)^2P\leq P_1=P_2=P.\label{P3}
\end{eqnarray}
As a result, three distinct interfaces are formed: 1–3, 2–3, and the original 1–2 interface, now modified by the presence of the surfactant. When the surfactant is formed, a nucleation is created with a small thickness layer. The condition for the nucleation is found
\begin{eqnarray}
\sqrt{2}\left(\frac{\xi_1}{\bar\xi_3}+\frac{\xi_2}{\bar\xi_3}\right)\left(\frac{\mu_3}{\bar\mu_3}\right)^{3/2}=\sqrt{K_{13}-\frac{\mu_3}{\bar\mu_3}}+\sqrt{K_{23}-\frac{\mu_3}{\bar\mu_3}}.\label{nucl}
\end{eqnarray}
The interfacial tension at the 1-2 interface with presence of the surfactant is defined as
\begin{eqnarray}
\label{gamma123}
    \gamma_{12(3)} &\equiv  &4 P \,\xi_2 \int _{-\infty }^{\infty} d \tilde z  \;\left [\left (\frac{\xi_1}{\xi_2}\frac{d\tilde \psi_1}{d\tilde z}\right )^2  
    +\left (\frac{d\tilde \psi_2}{d\tilde z }\right )^2  +    \left ( \frac{\bar\xi_3}{\xi_2}\frac{d\tilde \psi_3}{d\tilde z }\right )^2\right ],
\end{eqnarray}
whereas at two-phase $(i,j)$ coexistence, the interfacial tension are defined as \cite{Schaeybroeck2008,Barankov2002,Timmermans1998}
\begin{eqnarray}
\gamma_{ij}\equiv 4P\xi_2\int_{(i,-\infty)}^{(j,+\infty)}\left[\left(\frac{\xi_i}{\xi_2}\frac{d\tilde\psi_i}{d\tilde z}\right)^2+\left(\frac{\xi_j}{\xi_2}\frac{d\tilde\psi_j}{d\tilde z}\right)^2\right],\label{gammaij}
\end{eqnarray}
in which the notation $(i,-\infty)$ implies that component $i$ at $\tilde z\rightarrow-\infty$.

From an interfacial thermodynamics perspective, partial wetting corresponds to a competition between the excess free energies associated with three pairwise interfaces. In this regime, the interfacial tensions satisfy the inequality \cite{Rowlinson2002}
\begin{eqnarray}
\gamma_{12(3)}<\gamma_{13}+\gamma_{23},\label{antonov1}
\end{eqnarray}
indicating that the 1–2 interface is not completely “covered’’ by a microscopic film of component 3. The surfactant layer thickness is defined as $\tilde L=\tilde z^+-\tilde z^-$. A continuous wetting scenario arises when $\tilde L$ increases smoothly from zero to a macroscopic (divergent) value as the system parameters are varied. Along such trajectories in parameter space, the system evolves from partial wetting to complete wetting, where the interfacial free-energy balance reaches equality,
\begin{eqnarray}
\gamma_{12(3)}=\gamma_{13}+\gamma_{23}.\label{antonov11}
\end{eqnarray}
This condition eliminates the thermodynamic barrier to full adsorption of component 3 at the 1–2 interface, and the corresponding continuous transition is identified as a critical wetting phase transition. Within the GP mean-field framework, imposing the equalities $\tilde\psi_1=\tilde\psi_3$ at $\tilde z^-$ and $\tilde\psi_2=\tilde\psi_3$ at $\tilde z^+$ yields the critical wetting boundary at three-phase coexistence, which obeys \cite{Indekeu2025}
\begin{eqnarray}
\frac{\xi_1}{\xi_3\beta_{13}}+\frac{\xi_2}{\xi_3\beta_{23}}=\sqrt{2}.\label{critical}
\end{eqnarray}

A second possible scenario corresponds to a first-order wetting phase transition. In this regime, the interface free-energy balance prevents the formation of a microscopic film of component 3 at the 1–2 interface, and the third condensate remains metastable or unstable. The interfacial tensions must then satisfy Young’s law \cite{Rowlinson2002}
\begin{eqnarray}
\gamma_{12}\leq\gamma_{13}+\gamma_{23}.\label{Young}
\end{eqnarray}
which is the thermodynamic condition ensuring that the 1–2 interface cannot be completely wet by a surfactant layer of component 3. In geometric terms, Young’s law may be interpreted as the vanishing or reduction of the contact angle at the interface separating the three condensates, providing a microscopic analogue of the Young–Laplace relation in classical capillarity theory. As the system parameters are varied, a first-order wetting transition occurs when the inequality (\ref{Young}) becomes equality, eliminating the interfacial free-energy barrier and causing the film thickness to exhibit a discontinuous jump from zero to a finite value. This discontinuity distinguishes first-order wetting from the continuous (critical) scenario and reflects the metastability of the prewetting state near coexistence. At the two-phase coexistence regime, the interfacial tension  is evaluated using the definition (\ref{gammaij}) and the wave functions of condensates in the DPA \cite{Indekeu2015}, which yields
\begin{eqnarray}
\gamma_{ij}^{(\rm DPA)} &=& 2\sqrt{2} \frac{\beta_{ij}}{\sqrt{2}+\beta_{ij}}P (\xi_i + \xi_j),\label{gammaijDPA}
\end{eqnarray} 
where $\beta_{ij}=\sqrt{K_{ij}-1}$. Substituting Eq. (\ref{gammaij}) into Eq. (\ref{Young}), we obtain the following relation 
\begin{eqnarray}
\frac{\xi_1}{\xi_3}+\frac{\xi_2}{\xi_3}=\frac{\beta_{13}}{\sqrt{2}+\beta_{13}}\left(1+\frac{\xi_1}{\xi_3}\right)+\frac{\beta_{23}}{\sqrt{2}+\beta_{23}}\left(1+\frac{\xi_2}{\xi_3}\right).\label{first}
\end{eqnarray}
Equation (\ref{first}) represents the condition for the occurrence of a strongly first-order wetting transition, as discussed in Ref.\cite{Indekeu2025}.    

We now turn our attention to examining the conditions under which the interface between components 1 and 2 may become wetted by the third component. As outlined previously, the relative parameters defined in Eqs. (\ref{K}) and (\ref{xiratio}) are determined by intrinsic atomic parameters, particularly the intra- and interspecies scattering lengths. These interatomic interactions are experimentally tunable across several orders of magnitude in strength via the application of Feshbach resonances \cite{Inouye1998,Stan2004,Chin2010}. In the phase space spanned by these parameters, the wetting phase diagram at three-phase coexistence ($\mu_3/\bar\mu_3=1$) is determined by three boundaries: the nucleation line given by Eq. (\ref{nucl}), the critical and first-order wetting lines given by Eqs. (\ref{critical}) and (\ref{first}), respectively. Our system contains four independent parameters, $K_{13},K_{23},\xi_3/\xi_1$ and $\xi_3/\xi_2$. In Ref. \cite{Indekeu2025}, the intraspecies scattering lengths were held fixed so that the healing-length ratios defined in Eq. (\ref{xiratio}) remain constant, and the wetting phase transitions were investigated in the $(K_{13},K_{23})$-plane by varying the interspecies scattering lengths. Within this construction, the wetting phase diagram appears two degenerate points, which separate the phase diagram plane into three regions: one inner corresponds to the critical wetting phase transition and two outers correspond to the first-order wetting phase transition. 

We now investigate how the nucleation and wetting phase transitions are affected by variations in the healing-length ratios. Here, the relative coupling constants $K_{13}$ and $K_{23}$ are fixed, while the healing length ratios are modulated by varying the intraspecies scattering lengths. In particular, we hold $\xi_1,\xi_2$ constant, and investigate the effects of varying $\xi_3$.
\begin{figure}[t]
\centering
\includegraphics[width=0.6\textwidth]{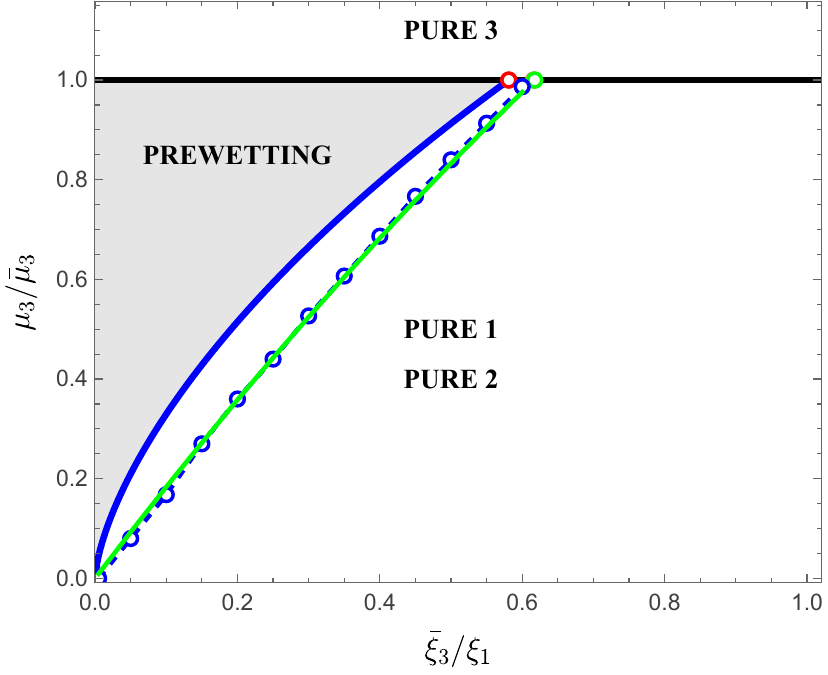}
\caption{The nucleation line of the surfactant as a function of $\bar\xi_3/\xi_1$ for $K_{13}=3,K_{23}=2K_{13}$ and $\bar\xi_3/\xi_2=2\bar\xi_3/\xi_1$.The blue solid line and the blue open circles connected by dashed lines represent the results obtained from the DPA and from numerical computations, respectively. The green solid line corresponds to that in Ref. \cite{Berx2026}.}\label{fig:nucleation}
\end{figure}
Fig. 1 shows the nucleation line within the DPA (blue solid line) in the $(\bar\xi_3/\xi_1,\mu_3/\bar\mu_3)$-plane for $K_{13}=3,K_{23}=2K_{13}$ and $\bar\xi_3/\xi_2=2\bar\xi_3/\xi_1$. The nucleation line intersects the three-phase coexistence line (black line) at  $\bar\xi_3/\xi_1=0.581$ (red dot). The corresponding results obtained from numerical computations are indicated by the blue open dots, which intersects the three-phase coexistence line (black line) at  $\bar\xi_3/\xi_1=0.616$ (pink dot). The corresponding analytical result in Ref. \cite{Berx2026} [see Eq. (45)] is shown by the green line, which intersects the three-phase coexistence line (green dot) at $\bar\xi_3/\xi_1=0.617$. At the three-phase coexistence point, the deviation of the DPA from the numerical result is approximately 5.66\%. This magnitude of deviation is comparable to that obtained for the two-component BECs absorbed at a hard-wall studied in Ref. \cite{DuyThanh2024}. The comparison demonstrates that the DPA provides a good description of the nucleation transition.

\begin{figure}[t]
\centering
\includegraphics[width=0.6\textwidth]{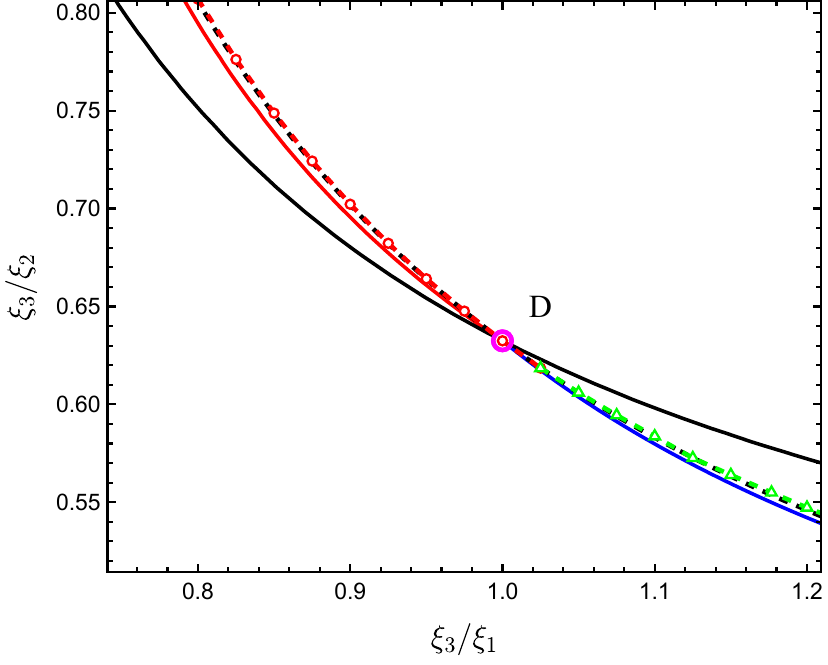}
\caption{Wetting phase diagram in the $(\xi_3/\xi_1,\xi_3/\xi_2)$-plane for $K_{13}=3$ and $K_{23}=2K_{13}$. The black solid line denotes the nucleation line, while the red and blue solid lines represent the first-order and critical wetting transition lines within the DPA, respectively. The black dashed line corresponds to the analytical result for the nucleation line in Ref. \cite{Berx2026}. Numerical results are indicated by symbols connected with dashed lines: red dots and green triangles for the first-order and critical wetting lines, respectively.}\label{fig:diagram1}
\end{figure}

As mentioned above, nontrivial wetting phase diagrams for the symmetric and asymmetric case were considered while the interspecies interaction strengths were varied. Here we investigate the wetting phase diagram in space of the intraspecies interaction strength.  The wetting phase diagram at the three-phase coexistence is sketched in Fig. \ref{fig:diagram1} at fixed $K_{13}=3$ and $K_{23}=2K_{13}$. The black, red and blue solid lines correspond to the nucleation line (\ref{nucl}), critical wetting line (\ref{critical}) and first-order wetting line (\ref{first}) within the DPA, respectively. A remarkable distinction between the present results and those reported in Ref. \cite{Indekeu2025} lies in the number of degenerate points observed. Specifically, in the current analysis, only a single degenerate point D is identified, whereas Ref. \cite{Indekeu2025} reported the existence of two degenerate points within the plane defined by the relative interspecies interaction coupling constants. This degenerate point corresponds to a first-order wetting transition without energy barrier, all surfactant layer thicknesses having the same value of the grand potential. Its uniqueness in the current parameter space implies a more constrained set of conditions under which the wetting transition occurs, reflecting the altered roles of the tunable parameters in this scenario. The reduction in the number of degenerate points highlights the sensitivity of the wetting behavior to the interplay between intra- and interspecies interactions, particularly under fixed relative coupling constants.

To evaluate the accuracy of the DPA, we perform numerical calculations for the phase diagram shown in Fig. \ref{fig:diagram1}. The nucleation line corresponding to the exact solution in Ref. \cite{Berx2026} is represented by the black dashed line. Firstly, the GP equations (\ref{coupledGP1}), (\ref{coupledGP2}) are solved numerically with the boundary conditions (\ref{boundaries2}) for given parameter values, with a numerical error of $10^{-5}$.
By numerically solving Eqs. (\ref{antonov11}) and (\ref{Young}), we obtain the points corresponding to the first-order and critical wetting transitions. In Fig. \ref{fig:diagram1}, these transition points are indicated by red circles and green squares, respectively. The results show that the nucleation, first-order, and critical wetting lines coincide, in contrast to the predictions of the DPA. This behavior is consistent with that observed in the plane of the relative coupling constants \cite{Berx2026}. It further indicates that the DPA is not suitable for describing the phase diagram in the regime of strong segregation between components 1 and 2 in the general case.

\begin{figure}[t]
\centering
\includegraphics[width=0.5\textwidth]{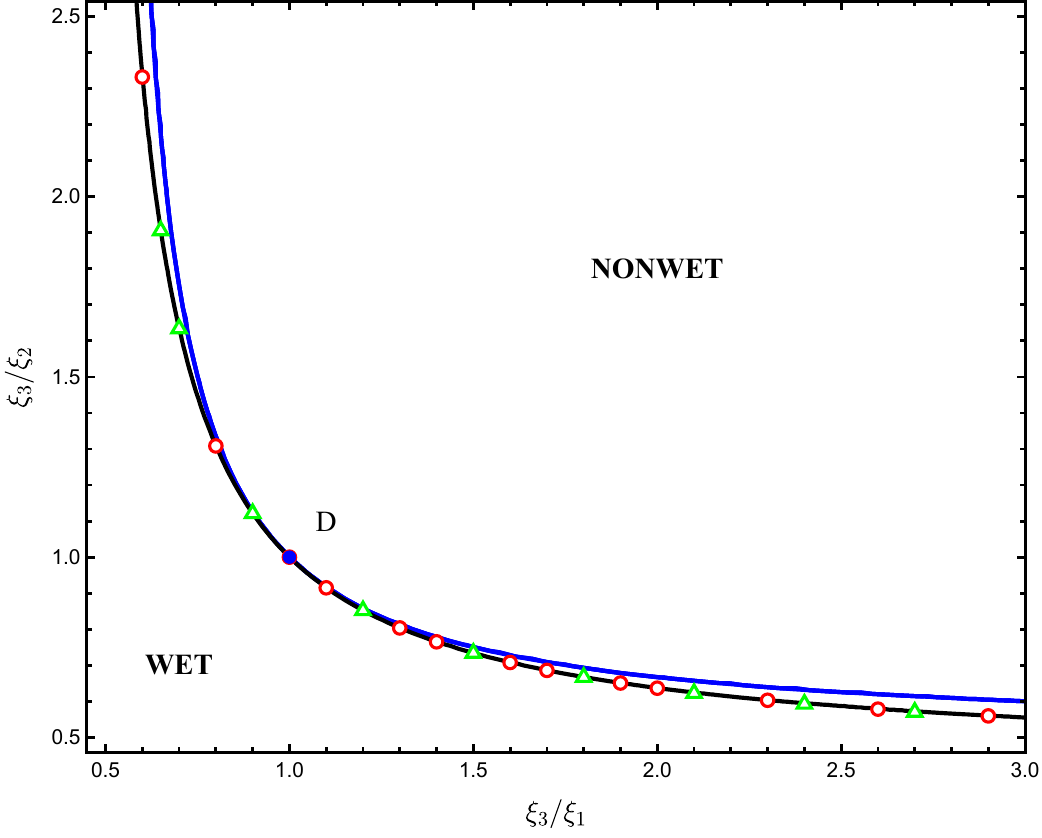}
\caption{Wetting phase diagram of the partial symmetric system in the $(\xi_3/\xi_1,\xi_3/\xi_2)$-plane at $K_{13}=K_{23}=3$. The blue solid line represents the phase boundary Eq. (\ref{symmetric1}) in the DPA whereas the black solid line expresses the analytical prediction obtained in Ref. \cite{Berx2026}. The numerical computations are shown by symbols: the red dots for the first-order wetting points and green triangles for the critical wetting points, respectively.}
\label{fig:symmetry}
\end{figure}
\begin{figure}[t]
\centering
\includegraphics[width=0.5\textwidth]{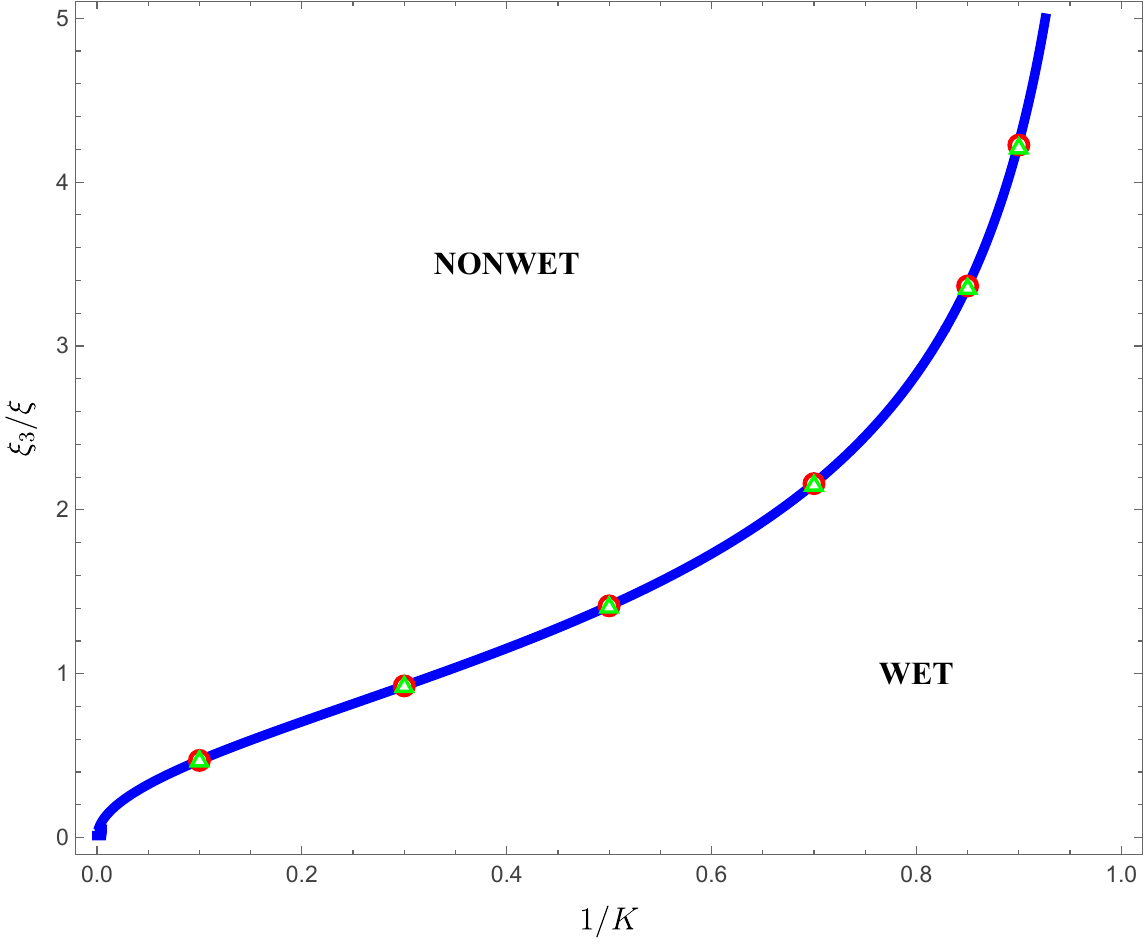}
\caption{Wetting phase diagram of the completely symmetric system in the $(1/K,\xi_3/\xi)$-plane. The blue solid line represents the phase boundary Eq. (\ref{sym}) in the DPA. The numerical computations are shown by symbols: the red dots for the first-order wetting points and green triangles for the critical wetting points, respectively.}
\label{fig:complete}
\end{figure}
Departing temporarily from the general case, we now consider several special cases in which the system exhibits symmetry. We first examine the partially symmetric system, in which the interspecies coupling constants are equal $K_{13}=K_{23}\equiv K$, or equivalently $\beta\equiv \beta_{13}=\beta_{23}=\sqrt{K-1}$. Within the DPA, the nucleation line (\ref{nucl}) reduces to
\begin{eqnarray}
\left(\frac{1}{\xi_3/\xi_1}+\frac{1}{\xi_3/\xi_2}\right)\left(\frac{\mu_3}{\bar\mu_3}\right)^{3/2}=\sqrt{2}\beta.\label{nucl1}
\end{eqnarray} 
At three-phase coexistence, the nucleation line (\ref{nucl1}), the first-order wetting line (\ref{first}) and the critical wetting line (\ref{critical}) coincide and reduce to
\begin{eqnarray}
\frac{1}{\xi_3/\xi_1}+\frac{1}{\xi_3/\xi_2}=\sqrt{2}\beta.\label{symmetric1}
\end{eqnarray}
This indicates that the phase boundary is degenerate, implying that the wetting transition is of strongly first-order character. This behavior is manifested by a discontinuous jump in the thickness of the surfactant layer from zero to a macroscopic value at the transition point \cite{Indekeu2025,Berx2026}, in agreement with the predictions of the GP theory and numerical calculations. In Fig. \ref{fig:symmetry}, we present the wetting phase diagram for the partially symmetric case with $K_{13}=K_{23}\equiv K= 3$ in the $(\xi_3/\xi_1,\xi_3/\xi_2)$-plane. The blue solid line corresponds to Eq. (\ref{symmetric1}) within the DPA, while the black solid line represents the GP nucleation line from Ref. \cite{Berx2026}. The red dots and green triangles denote the first-order and critical wetting transition points obtained from numerical computations, respectively. The results clearly show that the wetting transition is a degenerate first-order transition in both the DPA and GP theory. Interestingly, the DPA provides a good approximation in this symmetric case, in contrast to the general case investigated above.

For the complete symmetry, i.e., $K_{13}=K_{23}\equiv K$ and $\xi_1=\xi_2\equiv\xi$, the nucleation, first-order and critical wetting lines are described by the unique equation
\begin{eqnarray}
\frac{\sqrt{2}}{\xi_3/\xi}=\sqrt{K-1}.\label{sym}
\end{eqnarray}
Eq. (\ref{sym}) totally coincides with the one in the GP theory \cite{Berx2026} [see Eq. (31)]. The wetting phase diagram for the fully symmetric system is presented in Fig. \ref{fig:complete}. The phase boundary given by Eq. (\ref{sym}) is shown as the blue solid line. The wetting transition is of degenerate first-order character. Numerical results are indicated by red dots and green triangles corresponding to the first-order, and critical wetting lines, respectively. In excellent agreement with the predictions of the DPA, these three lines are found to coincide, providing further confirmation of the accuracy of the DPA. It is also again confirmed by point D in Fig. \ref{fig:symmetry} at which $\xi_1=\xi_2$ therefore two solid lines have a contact point.

\section{Conclusions\label{Conclusions}}

In this work, we have investigated the wetting phase transition in dilute ternary BECs. Motivated by experimental feasibility, we focused on scenarios in which the intraspecies atomic interactions are tunable, while the interspecies interactions are held fixed.

Our main findings are as follows:

\begin{itemize}

\item The nucleation transition has been analyzed for a specific set of intrinsic atomic parameters, providing insight into the onset of surfactant-mediated interfacial modification. Numerical calculations have been performed and compared with the predictions of the DPA, demonstrating that the DPA provides a reliable approximation. Although the GP theory yields analytical expressions, the DPA offers several advantages: it leads to simpler analytical form and yields only physically relevant solutions, whereas the GP theory may produce both physical and nonphysical solutions. 
    
\item The wetting phase diagram has been investigated in the plane of the healing-length ratios for the general case. Within the DPA, the wetting transition is predicted to be either first-order or critical. In contrast, the GP theory and numerical calculations yield a degenerate first-order wetting transition, characterized by the coincidence of the nucleation, first-order, and critical wetting lines. This discrepancy indicates that the DPA is not applicable in this case.

\item For the partially symmetric system, characterized by equal relative coupling constants, our calculations demonstrate that the DPA provides a reliable approximation for the study of wetting phase transitions, with the remarkable advantage of yielding simpler analytical expressions.

\item  In the completely symmetric system, defined by $\xi_1=\xi_2=\xi$ and $K_{13}=K_{23}=K$, the results from the DPA, GP theory and numerical calculations are in complete agreement. This implies that the DPA is a good tool to investigate the completely symmetric system.

\end{itemize}

These results contribute to a deeper understanding of interfacial phenomena in multicomponent BECs and provide theoretical guidance for experimental exploration of wetting transitions in ultracold atomic systems. The observation that the DPA provides an excellent description for symmetric systems, especially in the completely symmetric case, while it is not applicable to asymmetric systems, remains {\it an open question} that requires further investigation.

\section*{Acknowledgements}

We thank J. O. Indekeu and J. Berx for their discussion in the early stage of this work. This research is funded by Vietnam National Foundation for Science and Technology Development (NAFOSTED) under grant number 103.01-2023.12.

\end{document}